\documentclass
[aps,twocolumn,showpacs,groupedaddress,amsmath,floatfix,nofootinbib]{revtex4}

\advance\textheight by 0.25in
\advance\voffset by -0.125in

\usepackage{graphicx}
\usepackage{epsfig}
\usepackage{amsfonts}
\usepackage{amsmath,amssymb}

\newcommand{\be}{\begin{eqnarray}}
\newcommand{\ee}{\end{eqnarray}}
\newcommand{\non}{\nonumber\\}
 \newcommand{\ave}[1]{\langle {#1} \rangle}

 \newcommand{\GeV}{\hbox{GeV}}
 
 \newcommand{\Nch}{N_{\rm ch}}
 \newcommand{\tw}{\textwidth}

\begin{document}


\title{Detecting QGP with Charge Transfer Fluctuations}

\author{Sangyong Jeon}
\email[] {jeon@physics.mcgill.ca}
\affiliation{Physics Department, McGill University,
Montr{\'e}al, Canada H3A 2T8}
\affiliation{RIKEN-BNL Research Center, Upton NY 11973, USA}
\author{Lijun Shi}
\email[] {shil@physics.mcgill.ca}
\affiliation{Physics Department, McGill University,
Montr{\'e}al, Canada H3A 2T8}
\author{Marcus Bleicher}
\affiliation{Institut f\"{u}r Theoretische Physik,
 Johann Wolfgang Goethe -- Universit\"{a}t,
 Robert Mayer Str. 8-10,
 60054 Frankfurt am Main, Germany}
\date{\today}

\begin{abstract}

In this study, we analyze the recently proposed
charge transfer fluctuations within a finite pseudo-rapidity space.
As the charge transfer fluctuation is a measure of the local charge
correlation length,
it is capable of detecting inhomogeneity in the hot and dense
matter created by heavy ion collisions.
We predict that going from peripheral to central collisions,
the charge transfer fluctuations at midrapidity should decrease
substantially while
the charge transfer fluctuations at the
edges of the observation window should decrease by a small amount.
These are consequences of having a strongly inhomogeneous matter
where the QGP component is concentrated around midrapidity.
We also show how to constrain the values of the charge
correlations lengths in both the hadronic phase and the QGP phase using
the charge transfer fluctuations.

\end{abstract}

\maketitle

\section{Introduction \label{sect::intro}}

More than 20 years ago, Bjorken in his seminal paper\cite{Bjorken:1983AA}
considered the possibility that the central plateau around midrapidity
could be due to a hot and dense matter undergoing a boost-invariant expansion.
At high enough collision energies,
the temperature and density would be high enough for
the created matter to be composed of deconfined quarks and gluons
(quark-gluon plasma or QGP).

Recent studies
at RHIC have shown that extreme hot and dense matter has indeed
been created around midrapidity
with the energy density well above the expected transition density
\cite{Adcox:2004mh,Back:2004je,Adams:2005dq,Arsene:2004fa,Shuryak:2004cy}.
However, there are also evidences that the boost-invariance may not be a
feature of the created system even within the apparent plateau region.
For instance, the elliptic flow
measured by PHOBOS collaboration\cite{Back:2004mh}
shows no discernable plateau around the central (pseudo-)rapidity.

\begin{figure}[t]
\center
\includegraphics[width=0.45\tw,height=5.5cm,angle=0]{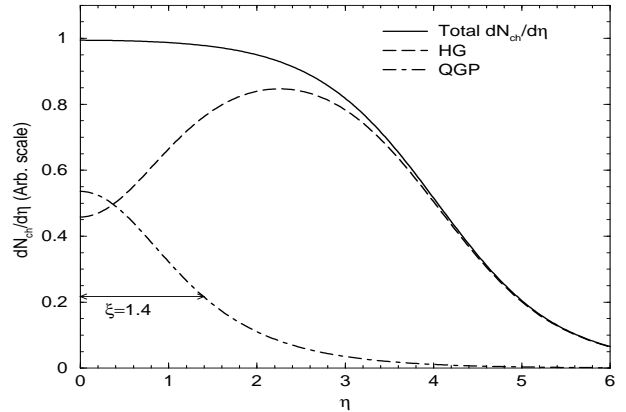}
\caption{
Illustration of a possible spatial inhomogeneity of the matter produced
in central heavy-ion reactions at RHIC.
The full line represent all charged particles
produced in the heavy-ion collisions.
The dot-dashed
line indicates the fraction of charged particles originating from a
QGP and the dashed
line is for the charged particles that never went through a
deconfined phase.
At midrapidity, the QGP concentration is high while
in the forward direction, almost all particles come from
the confined hadronic matter.
\label{fig:concept:inhomogeneous_matter}
}
\end{figure}

Taken together, the above evidences can be regarded as an indication
that the spatial extent
(in the pseudo-rapidity space) of the created QGP may be only a fraction of
the size of the plateau region.
For instance, the QGP component may be concentrated around the midrapidity
as schematically shown in Fig.\ref{fig:concept:inhomogeneous_matter}
while the rest of the system is mostly hadronic.

In view of such a possibility of having an inhomogeneous matter,
we should ask different questions about the produced matter at RHIC.
Namely, instead of asking whether we have created a QGP,
we should ask what fraction of the produced matter
went through the deconfined phase and how big was the size of the deconfined
phase.
If inhomogeneity is strong, these questions
should be answerable by some judicious choice of observables.

In a previous paper \cite{Shi:2005rc},
we have proposed a new observable, the charge transfer fluctuations, for
measuring the {\it local} charge correlation length.
Since the charge correlation lengths of a QGP and a hadronic gas can be
significantly different
\cite{Bass:2000az,Tonjes:2002us,Westfall:2004cq,Adams:2004kr,Jeon:2003gk},
the changes in charge transfer
fluctuations should signal the presence and the extent of the
inhomogeneity.
In Ref.\cite{Shi:2005rc}, we considered an ideal case assuming
nearly $4\pi$ detector with 100\,\% efficiency.
We argued that with such an ideal detector,
the charge transfer fluctuations should show a local minimum
where the QGP was formed.
In this followup paper,
we concentrate on a more realistic scenario.  The goal is to
predict what experiments at RHIC, STAR in particular, should observe.

The charge transfer fluctuation is defined
by \cite{Quigg:1973wy,Chao:1973jk}
  \begin{eqnarray}
  D_u(\eta) \equiv \langle u(\eta)^2 \rangle - \langle u(\eta)\rangle ^2.
  \label{eq:ChTr:chargetransfluc-def}
  \end{eqnarray}
The charge transfer $u(\eta)$ is in turn defined by the forward-backward charge
difference:
  \begin{eqnarray}
  u(\eta) &=& \left[ Q_F(\eta)-Q_B(\eta) \right] /2 \,,
  \label{eq:ChTr:fbcharge-asymmetry}
  \end{eqnarray}
where $Q_F(\eta)$ is the net charge in
the forward pseudo-rapidity (or rapidity) region of $\eta$
and $Q_B(\eta)$ is the net charge in the backward
pseudo-rapidity (or rapidity) region of $\eta$.
The fluctuation $D_u(\eta)$
is then a measure of the correlation between the
charges in the forward and the backward regions that are separated by the
cut located at $\eta$.
As the net charge is experimentally
easier to measure in the pseudo-rapidity space, from now on we will use the
pseudo-rapidity exclusively.  However, all formalism can be directly
translated to the rapidity.

\begin{figure}[t]
\centerline{
\includegraphics[width=0.5\tw]{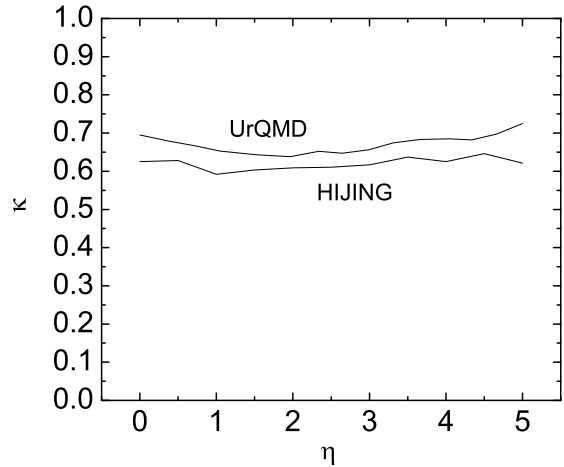}
}
\caption{
$\tilde{\kappa}(\eta) = (D_u(\eta)-\ave{\Delta Q^2}_{4\pi}/4)/(d\Nch/d\eta)$
generated by UrQMD Au-Au central collisions and HIJING central collisions.
The net charge fluctuation $\ave{\Delta Q^2}_{4\pi}$ of the produced
particles is in general non-zero
because the net charge of the spectators fluctuates.
Hence it must be subtracted from $D_u$ so that the ratio remains finite
near the beam rapidities.
}
\label{fig:UrQMD_Y}
\end{figure}

In Ref.\cite{Shi:2005rc},
we considered the case where
the experimental pseudo-rapidity coverage is much larger than the extent of
the QGP region.
Using a single-component and a two-component neutral cluster models,
we showed that the presence of a QGP component
results in a local minimum for $D_u(\eta)$ at the location of the highest
concentration of the QGP because it has a shorter charge correlation length.
The size of the dip can be then used to infer the size of the QGP region.
Data from $pp$ and $K^-p$ collisions in the energy ranges
of $p_{\rm lab} = 16\,\GeV$ to $p_{\rm lab} = 205\,\GeV$ show that the
charge transfer fluctuations is independent of rapidity
\cite{Chao:1973jk,Kafka:1975cz}.
This is also true for HIJING\cite{Wang:1991ht,Gyulassy:1994ew,Wang:1996yf} and
UrQMD\cite{Bass:1998ca,Bleicher:1999xi} events for Au-Au collisions
as shown in Fig.\ref{fig:UrQMD_Y}.

If the detector coverage is comparable or smaller than the QGP size,
then it is not likely that the local minimum can be observed.
Among the four experiments currently operating at RHIC, only the
STAR detector has enough coverage {\em and} charge-identification
capability to carry out the charge transfer fluctuation studies.
Still, the STAR pseudo-rapidity coverage $(|\eta| \le 1)$ is
comparable to the extent of the QGP region we estimated in
Ref.\cite{Shi:2005rc}.  Hence a quantitative analysis is necessary
to show the potential of the charge transfer fluctuation measurement.

The strength of the observed signal depends critically on the difference
between the charge correlation lengths in the QGP and the
hadronic phases.  One way to estimate the difference
is through the net charge fluctuations.
In Refs.\cite{Asakawa:2000wh,Jeon:2000wg}, two teams independently showed
that the net charge fluctuations per charged degrees of freedom,
$\ave{\Delta Q^2}/\ave{\Nch}$, can be 2 to 4 times smaller if the hadrons
came from a hadronizing QGP rather than from a hot resonance gas
\cite{Jeon:2003gk}.
Using neutral cluster models, one can show that this in turn implies
that the charge correlation length is
2 to 4 times smaller in a QGP than in a hadronic system.

\section{Thomas-Chao-Quigg Relationship and Non-QGP models}
\label{sect::TCQ}

Originally, Thomas and Quigg \cite{Quigg:1973wy} applied
the charge transfer fluctuations in the boost invariant case and obtained
\be
D_u = c\, \gamma \frac{\Nch}{Y}
\ee
where $\Nch/Y$ is
the value of the boost-invariant $d\Nch/dy$ and $\gamma$ is the
charge correlation length.  The proportionality constant $c$ depends on the
properties of the underlying clusters.  Later, Chao and Quigg generalized
this to smooth $d\Nch/dy$ cases and wrote down
the following Thomas-Chao-Quigg relationship \cite{Chao:1973jk}
\be
D_u(y) = \kappa\, \frac{d\Nch}{dy}
\label{eq:TCQ}
\ee
where $\kappa \propto \gamma$. They also showed that
$\kappa$ is constant (independent of $y$) for all
available elementary particle collision data at the time.

These original studies used a simple neutral cluster model where
an underlying cluster with the rapidity $y$ decays into 2 charged particles
and a single neutral particle.
The rapidities of these 3 decay products are given by
$(y{-}\Delta, y, y{+}\Delta)$ with $\Delta \propto \gamma$.
In Ref.\cite{Shi:2005rc}, we generalized this simple model so that the
the joint probability density for the charged decay products is given by
(switching to pseudo-rapidity)
\be
f(\eta^+, \eta^-) = R(r)F(Y)
\label{eq:f_sep}
\ee
where $r=\eta^+{-}\eta^-$ and $Y=(\eta^+{+}\eta^-)/2$.
Here the function $F(Y)$ can be interpreted as the rapidity distribution of
the clusters and $R(r)$ can be interpreted as the rapidity distribution of
the decay products given the cluster rapidity $Y$.
We then showed that the above
Thomas-Chao-Quigg relationship (\ref{eq:TCQ}) with a constant $\kappa$
is exactly satisfied
if
\be
R(r) = \frac{1}{2\gamma}\exp(-|r|/\gamma)
\label{eq:Rr}
\ee
with $\gamma = 2\kappa$ while $F(Y)$ is chosen so that
the single particle distribution
$g(\eta) = \int_{-\infty}^\infty d\eta'\, f(\eta',\eta)$
yields the normalized $d\Nch/d\eta$.
Realizing that the above $R(r)$ is
the Green function of the operator $d^2/dr^2 - 1/\gamma^2$, one obtains
\be
2\ave{M_0} F(\eta) =
\left(1 - \frac{\gamma^2}{4}
\frac{d^2}{d\eta^2}\right) \frac{d\Nch}{d\eta}
\label{eq:Feta}
\ee
where $\ave{M_0}$ is the average number of the neutral clusters.
The charge density $d\Nch/d\eta$ is modelled with a Wood-Saxon
form in this study and more sophisticated fittings are also
possible\cite{Jeon:2003nx}.

As shown in Fig.\ref{fig:UrQMD_Y}, non-QGP
models of heavy ion collisions also satisfy the Thomas-Chao-Quigg
relationship with a constant $\kappa \approx 0.6 - 0.7$ or equivalently
$\gamma \approx 1.2 - 1.4$.

One should emphasize here that the Thomas-Chao-Quigg relationship is for the
case where the {\em whole} (pseudo-) rapidity space is observed.
If the observational window is limited, then the ratio
\be
\bar{\kappa}(\eta) = \frac{\bar{D}_u(\eta)}{d\Nch/d\eta}
\ee
will no longer be independent of $\eta$ even if $\gamma$ is constant.
The bar over $\kappa$
and $D_u$ indicates that they are measured only within a finite observation
window.
In Ref.\cite{Shi:2005rc} we showed that
when the observation window is confined
to $|\eta| \le \eta_o$, the charge transfer fluctuation is
\be
\bar{D}_u(\eta)
=
\frac{\ave{\Delta Q^2}}{4}
+
2\ave{M_0}
\int_{-\eta_o}^\eta d\eta^-\int_{\eta}^{\eta_o} d\eta^+
f(\eta^+, \eta^-)
\label{eq:barDu}
\ee
where
\be
\ave{\Delta Q^2}
=
4\ave{M_0}\int_{-\infty}^{-\eta_o}d\eta^+\int_{-\eta_o}^{\eta_o}d\eta^-
f(\eta', \eta)
\ee
is the
net charge fluctuation within $|\eta| < \eta_o$ and $M_0$ is the total
number of the neutral clusters.

\begin{figure}[t]
\centerline{
\includegraphics[width=0.5\tw]{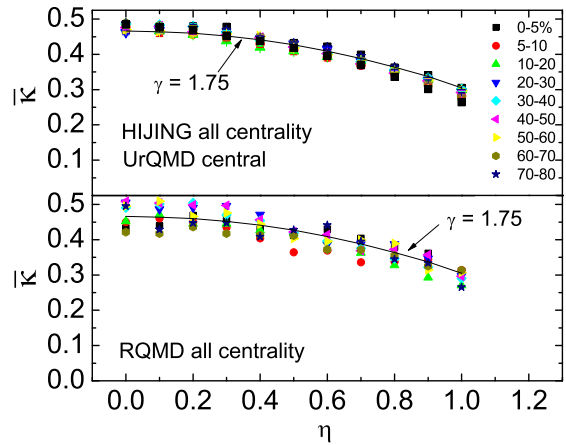}}
\caption{The charge transfer fluctuation
is shown as a function of pseudo-rapidity
$\eta$.
Different symbols represent standard centralities as shown in the symbol
legend.
These results are from $8,400$ RQMD events for
$\sqrt{s}=130\,\GeV$ Au-Au collisions and $50,000$ HIJING events at
$\sqrt{s}=200\,\GeV$.
STAR acceptance and efficiency
are taken into account in a simple way by restricting $p_T >
0.1\,\GeV$ and also randomly taking out 10\,\% of the charged particles.
}
\label{fig:HIJING}
\end{figure}

If $d\Nch/d\eta$ is constant
within the window as is the
case for the STAR Au-Au data at RHIC energies,
it is clear that $\bar{\kappa}(\eta)$ still
varies with $\eta$ even if $\kappa$ itself is constant
and such variation is entirely
given by the second term in Eq.(\ref{eq:barDu}).
In Fig.\ref{fig:HIJING}, we show HIJING, UrQMD and
RQMD\cite{Sorge:1992ej,Sorge:1995dp} results at various
centralities together with a single
component model fit.
The calculation is done with the STAR acceptance
$\eta_o = 1$ and $p_T > 0.1\,\GeV$.
Also the STAR detection
efficiency is taken into account in a simple way by
either randomly taking out
10\,\% of charged particles (for simulations) or
by adding 10\,\% of uncorrelated
charged particles (for the neutral cluster model).
>From these figures, it is clear that the non-QGP model results are
very well described by a single-component neutral cluster model with
$\gamma \approx 1.75$ \emph{independent of centralities}.
The discrepancy between this value of $\gamma$ and the $\gamma$ obtained in
the full phase space study is partly due to the acceptance cuts and partly
due to the presence of large $\ave{\Delta Q^2}$ within $|\eta|\le \eta_o$.
In the present case of limited observational window, the net
charge fluctuations should {\em not} be subtracted from $\bar{D}_u$.

\section{A QGP Model -- Central Collisions}

In RHIC energy heavy-ion reactions, the energy density of the created
systems vary with the centrality of the collisions.
The results of HIJING simulations and its single-component
fit should correspond to the peripheral collision results.
In central collisions, one expects that a QGP is formed
and is concentrated around midrapidity.
Hence, the final state particles can have 2 different origins in central
collisions: Some particles will
come from the hadronized QGP and
others will come from the non-QGP hadronic
component of the system.

In Ref.\cite{Shi:2005rc} we used a two-component neutral cluster model
and showed that even for such an inhomogeneous matter Eq.(\ref{eq:barDu})
still holds if one substitutes the $f(\eta^+,\eta^-)$ with
the following combination of
the QGP correlation function $f_{\rm QGP}$
and the hadronic gas correlation function $f_{\rm HG}$
\be
f(\eta^+, \eta^-) =
(1-p) f_{\rm HG}(\eta^+, \eta^-)
+
p f_{\rm QGP}(\eta^+, \eta^-)
\ee
Here $p$ is the fraction of the charged particles
originating from the QGP component in the whole phase space.

Each charge correlation function is assumed to have the
separable form as in Eqs.(\ref{eq:f_sep}) and (\ref{eq:Rr}).
In the rest of this manuscript, we denote the hadronic correlation length
with $\gamma_1$ and the QGP correlation length with $\gamma_2$.
The charge correlation length is expected to be a factor
of $2-4$ times smaller in
the QGP than in the hadronic matter\cite{Jeon:2003gk},
hence, $\gamma_2 < \gamma_1$.
The $F(Y)$ functions for each components are chosen as follows.
\be
(1-p)F_{\rm HG}(Y)
& =& W(Y) - p\,g_1(Y)
\\
p F_{\rm QGP}(Y)
& =& p\, g_2(Y)
\ee
The functions $g_1$ and $g_2$ are required to satisfy
\be
\rho(\eta)
& = &
\int_{-\infty}^\infty 
\frac{d\eta'}{2\gamma_1} \, g_1((\eta'+\eta)/2)\,
e^{-|\eta'-\eta|/\gamma_1}
\non
& = &
\int_{-\infty}^\infty \frac{d\eta'}{2\gamma_2} \, g_2((\eta'+\eta)/2)\,
e^{-|\eta'-\eta|/\gamma_2}
\label{eq:rhoeta}
\ee
The function $W(Y)$ is again given by
the left hand side of Eq.(\ref{eq:Feta}) with $\gamma = \gamma_1$.
The function $\rho(\eta)$ has the meaning of
the normalized $d\Nch/d\eta$ for the hadrons coming from the QGP component
(the dot-dashed line in Fig.\ref{fig:concept:inhomogeneous_matter}).

For $g_1(Y)$ we chose a gaussian
\be
g_1(Y) = \frac{1}{\sqrt{2\pi \sigma_1^2}}\,e^{-Y^2/2\sigma_1^2}
\ee
which fixes $g_2(\eta)$ to be
\be
\lefteqn{g_2(\eta) =
\frac{\gamma_2^2}{\gamma_1^2} g_1(\eta)} &&
\non
& & +
\left(1 - \frac{\gamma_2^2}{\gamma_1^2} \right)
\int_{-\infty}^\infty 
\frac{d\eta'}{2\gamma_1} \, g_1((\eta'+\eta)/2)\,
e^{-|\eta'-\eta|/\gamma_1}
\ee
again using the fact that $R_{\rm HG}$ and $R_{\rm QPG}$ are Green
functions.
The resulting charged particle distribution typically looks like
Fig.\ref{fig:concept:inhomogeneous_matter}.

So far we have introduced
4 parameters, $p, \gamma_1, \gamma_2$ and $\sigma_1$.
We can fix
\be
\gamma_1 = 1.75
\ee
by assuming that the HIJING data shown
in Fig.~\ref{fig:HIJING} is consistent with peripheral collisions at
RHIC\footnote{The lines shown in Fig.~\ref{fig:HIJING} are consistent
with $1.5 \lesssim \gamma_1 \lesssim 2.0$. The value $\gamma_1 = 1.75$
provides a good fit the whole set.  The analysis given here can be
repeated with any values within this range with minimal changes.}.%
For central collisions,
we further fix $p$ by requiring that
it is the maximum possible value that satisfies
$F_{\rm HG}(Y) \ge 0$. This in practice means $p = W(0)/g_1(0)$.
For peripheral collisions, the QGP fraction $p$ is presumed to be zero.

The remaining 2 parameters $\gamma_2$ and $\sigma_1$ can be further
constrained by considering the experimental data
on the net charge fluctuations from
STAR~\cite{Adams:2003st}.
>From Ref.\cite{Adams:2003st}, we can infer that for central collisions
\be
\bar{\kappa}(\eta_o=1.0)
\approx 0.27
\ee

\begin{figure}[Ht]
\center\includegraphics[width=0.45\tw]{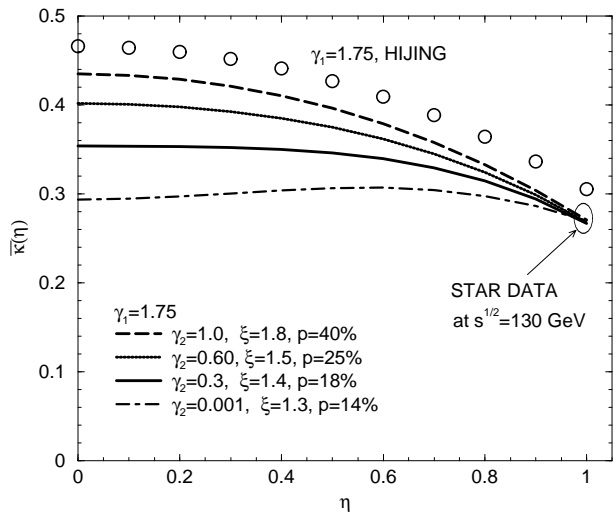}
\caption{
The results of two-component model calculations together with the
single-component result that describes the HIJING data.
Also shown is the position of the data point deduced from the STAR net
charge fluctuations measurement.}
\label{fig:QGP}
\end{figure}

The results of numerically exploring $0 < \gamma_2 \le 1.0$
are shown in Fig.~\ref{fig:QGP}.
At each $\gamma_2$, there is a single $\sigma_1$
that makes $\bar{\kappa}(1.0) = 0.27$.  With $\sigma_1$ thus fixed, one can
then calculate the extent of the QGP component in the pseudo-rapidity space
by calculating $p$ and the rms-width
\be
\xi = \sqrt{\sigma_1^2 + \gamma_1^2/2}
\ee
of $\rho(\eta)$ (c.f.Eq.(\ref{eq:rhoeta})).
Note that $\xi > \gamma_1/\sqrt{2} = 1.24$ is always larger than
the STAR rapidity window size $\eta_o$.

The results of the charge transfer fluctuation calculations
in the two-component model are shown
Fig.~\ref{fig:QGP}. For comparison, we have also plotted
results of the single-component model calculations
in Fig.~\ref{fig:singles}.
The results of
the single-component and two-component model calculations
are distinctive enough that
the measurement of $\bar{\kappa}(\eta)$ can clearly differentiate the two
model scenarios, hence providing a way to show the existence of the QGP.

For the two-component model, we find the single data point at
$\eta=1.0$ does not constrain the overall shape of $\bar{\kappa}(\eta)$
that much.
We can find a range of
possible parameter sets that give
the same $\bar{\kappa}(1.0)=0.27$ while the
corresponding shapes of $\bar{\kappa}(\eta)$ are all very different.
The biggest difference
among these parameter sets is the value of $\bar{\kappa}(\eta)$ at
$\eta = 0.0$, where the concentration
of the QGP component is strongest.
The four lines with $0.001 \le \gamma_2 \le 1.0$ in
Fig.~\ref{fig:QGP} represent typical results in two-component
model calculations.

In a clear contrast, in the single component model
the value of $\bar{\kappa}(1.0)$ completely fixes the shape of
$\bar{\kappa}(\eta)$ in the entire interval $0.0\le \eta \le 1.0$.
To have $\bar{\kappa}(1.0)=0.27$,
the correlation length must be $\gamma=1.3$.
As shown in Ref.\cite{Shi:2005rc},
$\bar{\kappa}(\eta)$ is proportional to $\gamma$ in the limit $\gamma/\eta_o
\ll 1$.  Hence reduction in $\gamma$ results in the overall reduction of the
$\bar{\kappa}(\eta)$ in the whole range.

>From these considerations, we can say that
the measurement of the charge transfer fluctuations in
the entire range $0.0<\eta <1.0$ for various centralities
will be a critical test for the existence of QGP.
If the cental collision data shows a clear reduction
from the $\gamma=1.3$ line in Fig.~\ref{fig:singles},
it can only be explained by the presence of the second component.
We also expect the amount of QGP would grow as one goes
from peripheral to central collisions.  Therefore, the most
peripheral collision
data for $\bar{\kappa}(\eta)$ will behave more like the single-component
results while the most central collision data for $\bar{\kappa}(\eta)$ will
behave more like the two-component results.
We predict that the most central collision data should lie between
the two solid lines with $\gamma_2=0.6$ and $0.3$ in Fig.\ref{fig:QGP}.
For comparison, we have also shown two extreme cases with very
large
and very small QGP charge correlation lengthes where $\gamma_2=1.0$ and
$0.001$.

In summary, we have made the following prediction for the charge transfer
fluctuations within limited pseudo-rapidity range of $|\eta|\le 1.0$.
Suppose that the systems created by two colliding heavy ions at RHIC are
inhomogeneous mixtures of the QGP component and the non-QGP hadronic
component.  Further suppose that the QGP component is concentrated near
midrapidity and as one goes from peripheral to central collisions,
the amount of the created QGP increases.
Then the charge transfer fluctuations,
$\bar{\kappa}(\eta) = \bar{D}_u(\eta)/(d\Nch/d\eta)$,
should show the following signature of the presence of the QGP:
Going from peripheral to central collisions,
$\bar{\kappa}(\eta=0.0)$ should decrease
substantially from about 0.45 to 0.35 while $\bar{\kappa}(\eta=1.0)$
should decrease by a small amount from about 0.3 to 0.27.
>From the amount of the reductions, we can then infer the size and the
fraction of the QGP matter.
In contrast, if a QGP is not formed at any centralities, then the data
points from different centralities should all fall on the
same curve.

\section{Uncorrelated charges \label{sect::UNCORR}}

The detection efficiency for charged particles is typically less than
100\,\%
in real experiments. This has been a concern in the measurement of the net
charge fluctuations. In the previous paper, we have argued that this effect
is small and will not effect the qualitative arguments we had
\cite{Shi:2005rc}. However, with limited observation window, the difference
between one and two component model is more quantitative and the detector
efficiency deserves some attention.

In terms of the pair correlations, the non-ideal detector efficiency
renders some of the correlated pair to be uncorrelated.
The relevant formulas are already worked out in
the previous paper \cite{Shi:2005rc}.

\begin{figure}[t]
\center\includegraphics[width=0.45\tw]{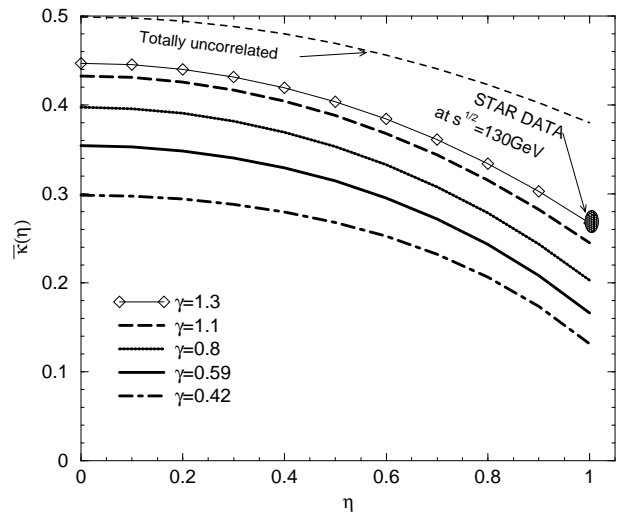}
\caption{
Single-component results.
The lower four lines
have the same $\bar{\kappa}(1)$ values as
the four two-component model results in Fig.\ref{fig:QGP}.
Also shown are the results for the totally uncorrelated case (upper most line)
and the single-component model result that has $\bar{\kappa}(1) = 0.27$
(thin line connecting the diamond symbols).
}
\label{fig:singles}
\end{figure}

If the detector efficiency is $\epsilon$ and
all the charged particles are correlated,
then $1-\epsilon$ of detected particles
will become uncorrelated because their partners are not detected.
The corresponding charge transfer fluctuations are
$\bar{\kappa} = \epsilon\bar{\kappa}_{\rm corr}
+ (1-\epsilon)\bar{\kappa}_{\rm uncorr}$,
where $\bar{\kappa}_{\rm corr}$ is
for the fully correlated charged particles and
$\bar{\kappa}_{\rm uncorr}$ is for the uncorrelated charged particles.

The shape of $\bar{\kappa}_{\rm uncorr}$ is shown in Fig.\ref{fig:singles}.
The shape is flatter than the correlated cases and it is always above
$\bar{\kappa}_{\rm corr}$.
The detector efficiency in STAR experiment
is about $85 - 90\,\%$.
Having $10\,\%$ of uncorrelated charges will
slightly increase $\bar{\kappa}(\eta)$ and
make the overall shape a little bit flatter.
This, however, will not change
much of the results for $\kappa(\eta)$.
The signature for the appearance of the second QGP-like
component is still present, and we can still put an upper limit on the
$\gamma_{\rm QGP}$ if a significant reduction of $\kappa(0.0)$ is observed.
All the results presented in this paper already considered the effect of
uncorrelated charged particles.

\section{Summary \label{sect::summary}}

In this paper, we proposed
the charge transfer fluctuation as a good observable capable of detecting the
{\em local} presence of a QGP in a limited pseudo-rapidity space.
In contrast, the net charge fluctuations\cite{Asakawa:2000wh,Jeon:2000wg}
and the width of the balance function\cite{Bass:2000az} are only
sensitive to the presence of a QGP when
the fraction of the QGP component is substantial
in the whole observational window.
Since longitudinal inhomogeneity is
expected from both theoretical considerations and experimental observations,
it is important to have an observable that is sensitive to it.
Furthermore, such inhomogeneity may explain why the net charge fluctuations
did not show a strong signal even though the underlying net charge
fluctuations could be strongly reduced in the QGP phase.

In this study, we showed that the
three hadronic models, HIJING, RQMD and UrQMD are consistent with a
single-component model with fixed charge correlation length of about about
$\gamma=1.75$.
If a QGP is created in central heavy ion collisions, a significant deviation
from this behavior is expected in the data.  Specifically, if the QGP
component is concentrated around the midrapidity and tapers off
going away from midrapidity,
then one should see the following trends in the data:

\begin{enumerate}
\item[(i)] The overall values of $\bar{\kappa}(\eta)$ must decrease going from
peripheral collisions to central collisions.  This indicates that more
QGP is being created.

\item[(ii)] The value of $\bar{\kappa}(1.0)$ should change moderately from
around 0.30 to 0.27 from peripheral to central collisions.
These values correspond to
the HIJING, UrQMD and RQMD value for peripheral Au-Au collisions,
and the measured value of
$\ave{\Delta Q^2}/\ave{\Nch}$ for central Au-Au collision
from STAR. This small reduction indicates that near $\eta = \pm 1$,
the contribution from the QGP component is already much reduced.

\item[(iii)] The reduction in the value of $\bar{\kappa}(0.0)$ should be
larger than the reduction in the value of $\bar{\kappa}(1.0)$ as the QGP
component is more concentrated around midrapidity.
Going from peripheral to central events,
the value of $\bar{\kappa}(0.0)$ should vary from
around 0.45 (HIJING, UrQMD, RQMD) down to 0.35.
The value of $\bar{\kappa}(0.0)$ puts a severe constraint on the value of
the charge correlation length for the QGP component.
\end{enumerate}

The change in the value of $\bar{\kappa}(0.0)$ may not seem large. But keep
in mind that the value of $\bar{\kappa}(1.0)$ cannot be lower than the
already measured value of $0.27$.  Hence $\bar{\kappa}(0.0)$, too, cannot go
lower than that.
If $\bar{\kappa}(0.0)$ is measured to be close to 0.35, it is impossible
to explain this without the presence of
the second phase with a very short charge correlation length.

In summary, we propose that the charge transfer fluctuation is a sensitive
observable to find the presence and extent of the QGP created in high energy
heavy ion collisions.  In addition, this observable is relatively easy to
measure and does not require the net charge conservation correction.
We strongly suggest the experimental group to measure the charge transfer
fluctuations.

\begin{acknowledgments}

The authors thank J.Barrette for his useful suggestions. 
We also thank V.Topor Pop for providing us with the RQMD data and his help
in running the HIJING code.
The authors are supported in part by the Natural Sciences and
Engineering Research Council of Canada and by le Fonds
Nature et Technologies of Qu\'ebec.
S.J.~also
thanks RIKEN BNL Center and U.S. Department of Energy
[DE-AC02-98CH10886] for
providing facilities essential for the completion of this work.

\end{acknowledgments}

\end{document}